\documentstyle[11pt]{article}
\let\a=\alpha

\let\la=\label  
   
 \def\bd{\begin{document}} \def\ed{\end{document}}
\def\ds{\documentstyle} \let\fr=\frac \let\bl=\bigl \let\br=\bigr
\let\Br=\Bigr \let\Bl=\Bigl 
\let\bm=\bibitem
\let\na=\nabla
\let\pa=\partial \let\ov=\overline 
\newcommand{\be}{\begin{equation}} 
\newcommand{\ee}{\end{equation}} 
\def\ba{\begin{array}}
\def\ea{\end{array}}
\def\ft#1#2{{\textstyle{{\scriptstyle #1}\over {\scriptstyle #2}}}}
\def\fft#1#2{{#1 \over #2}}
\def\del{\partial}
\def\sst#1{{\scriptscriptstyle #1}}
\def\oneone{\rlap 1\mkern4mu{\rm l}}
\def\td{\tilde}
\def\wtd{\widetilde}
\newcommand{\ho}[1]{$\, ^{#1}$}
\newcommand{\hoch}[1]{$\, ^{#1}$}
\newcommand{\bea}{\begin{eqnarray}} 
\newcommand{\eea}{\end{eqnarray}} 
\newcommand{\ra}{\rightarrow}
\newcommand{\lra}{\longrightarrow}
\newcommand{\Lra}{\Leftrightarrow}
\newcommand{\ap}{\alpha^\prime}
\newcommand{\bp}{\tilde \beta^\prime}
\newcommand{\tr}{{\rm tr} }
\newcommand{\Tr}{{\rm Tr} } 
\newcommand{\NP}{Nucl. Phys. }
\newcommand{\tamphys}{\it Center for Theoretical Physics\\
Texas A\&M University, College Station, Texas 77843}
\newcommand{\auth}{M. J. Duff\hoch{\dagger}, H.
L\"u\hoch{\ddagger} and C. N. Pope\hoch{\ddagger}}

\def\kpfp{(v_\a\, e^{\phi} + \td v_\a\, e^{-\phi})}
\def\kpfm{(v_\a\, e^{\phi} - \td v_\a\, e^{-\phi})}

\begin{document}

\hfill{CTP-TAMU-9/96}

\hfill{hep-th/9603037}

\vspace{20pt}

\begin{center}
{ \large {\bf HETEROTIC PHASE TRANSITIONS AND SINGULARITIES OF THE 
GAUGE DYONIC STRING}}

\vspace{30pt}

\auth

\vspace{15pt}

{\tamphys}

\vspace{40pt}

\underline{ABSTRACT}
\end{center}

Heterotic strings on $R^6 \times K3$ generically appear to undergo some
interesting new phase transition at that value of the string coupling for
which the one of the six-dimensional gauge field kinetic energies changes
sign.  An exception is the $E_8 \times E_8$ string with equal instanton
numbers in the two $E_8$'s, which admits a heterotic/heterotic self-duality.
In this paper, we generalize the dyonic string solution of the
six-dimensional heterotic string to include non-trivial gauge field
configurations corresponding to self-dual Yang-Mills instantons in the four
transverse dimensions.  We find that vacua which undergo a phase transition
always admit a string solution exhibiting a naked singularity, whereas for
vacua admitting a self-duality the solution is always regular.  When there
is a phase transition, there exists a choice of instanton numbers for which
the dyonic string is tensionless and quasi-anti-self-dual at that critical
value of the coupling.  For an infinite subset of the other choices of
instanton number, the string will also be tensionless, but all at larger
values of the coupling.

{\vfill\leftline{}\vfill
\vskip	10pt
\footnoterule
{\footnotesize
	\hoch{\dagger}	Research supported in part by NSF Grant	PHY-9411543
\vskip	-12pt} \vskip	10pt {\footnotesize
	\hoch{\ddagger}	Research supported in part by DOE 
Grant DE-FG05-91-ER40633 \vskip	-12pt}}

\pagebreak
\setcounter{page}{1}

%\section{Introduction}
%\la{Introduction}

The purpose of this paper is to connect two hitherto unrelated phenomena
appearing in six-dimensional heterotic string theory: phase transitions
\cite{evidence,mv,afiq} and the dyonic string soliton \cite{Rahmfeld2}.
First we extend the {\it neutral} dyonic string solution to the {\it gauge}
solution by including non-trivial gauge field configurations corresponding
to self-dual Yang-Mills instantons in the four dimensions transverse to the
string.  This is similar to the way that the gauge fivebrane
\cite{Strominger} is related to  the neutral fivebrane \cite{elementary,chs}
in ten dimensions.  Then we show that vacua which undergo a phase transition
always admit a gauge dyonic string solution exhibiting a naked singularity,
whereas for vacua admitting a heterotic/heterotic self-duality
\cite{evidence} the solution is always regular. We find that when there is a
phase transition, there exists a choice of instanton numbers for which the
dyonic string is tensionless and quasi-anti-self-dual at that critical value
of the coupling.  For an infinite subset of the other choices of instanton
number, the string will also be tensionless, but all at larger values of
the coupling. 

      Let us begin by recalling the evidence for phase transitions in the
six-dimensional heterotic string \cite{evidence}.  Before the recent
interest in a duality between heterotic and Type IIA strings, it was
conjectured that in $D\leq6$ dimensions there ought to exist a duality
between one heterotic string and another \cite{Khuristring}. A comparison of
the (purely ``electric'') fundamental string solution \cite{Dabholkar} and
the (purely ``magnetic'') dual solitonic string solution
\cite{Lublack,Minasian} suggests the following $D=6$ duality dictionary: the
dilaton $\tilde \phi$, the canonical metric $\tilde g_{\sst{MN}}$ and
$3$-form field strength $\tilde H$ of the dual string are related to those
of the fundamental string,  $\phi$, $g_{\sst{MN}}$ and $H$ by the
replacements $\phi \rightarrow \tilde \phi=-\phi$, $g_{\sst{MN}} \rightarrow
\tilde g_{\sst{MN}}=g_{\sst{MN}}$, $H \rightarrow \tilde H=e^{-2\phi}\,
{}^*\! H$, where ${}^*$ denotes the Hodge dual. In going from the
fundamental string to the dual string, one also interchanges the roles of
worldsheet and spacetime loop expansions. Moreover, since the dilaton enters
the dual string equations with the opposite sign to the fundamental string,
the strong coupling regime of the string should correspond to the weak
coupling regime of the dual string \cite{Lublack,Minasian}:  ${\lambda}_6 =
\,<\!e^{\phi}\!>\, =1/{\tilde{\lambda}_6}$ where $\lambda_6$ and $\tilde
\lambda_6$ are the fundamental string and dual string coupling constants.
Because this duality interchanges worldsheet and spacetime loop expansions,
it exchanges the tree level Chern-Simons contributions to the Bianchi
identity 
\be 
d H = \fft{\a'}{4}(-\tr\, R\wedge R + \sum_\a v_\a\tr\, F_\a\wedge F_\a)\ ,
\la{Chern}
\ee
with the one-loop Green-Schwarz corrections to the field equations
\be
d \wtd H =  \fft{\a'}{4}(-\tr\, R\wedge R + \sum_\a \td v_\a\tr\, F_\a\wedge
F_\a)\ .
\la{Green}
\ee
Here $F_\alpha$ is the field strength of the $\alpha$'th component of the
gauge group, $\tr$ denotes the trace in the fundamental representation, and
$v_\alpha,\tilde v_\alpha$ are constants.  (As explained in \cite{evidence},
we may, without loss of generality, choose the string tension measured in
the string metric and the dual string tension measured in the dual string
metric to be equal.) In fact, the Green-Schwarz anomaly cancellation
mechanism in six dimensions requires that the anomaly eight-form $I_8$
factorize as a product of four-forms, $I_8 \sim dH \wedge d\tilde H$, and a
six-dimensional string-string duality with the general features summarized
above would exchange the two factors. 

  To see where the phase transition makes its appearance, let us recall that
in \cite{Sagnotti} corrections to the Bianchi identities of the type
(\ref{Chern}) and to the field equations of the type (\ref{Green}) were
shown to be entirely consistent with supersymmetry, with no restrictions on
the constants $v_\alpha$ and $\tilde v_\alpha$.  Moreover, supersymmetry
relates these coefficients to the gauge field kinetic energy.  In the
canonical metric, the dilaton dependence of the kinetic energy of the
gauge field $F_\alpha{}_{MN}$ is 
\be 
L_{\rm gauge} \sim \sqrt{-g}\sum_\alpha \left( v_\alpha
e^{-\phi}  +\tilde v_\alpha e^{\phi} \right)\tr
F_\alpha{}_{MN}F_\alpha{}^{MN}\ . 
\la{happiness}
\ee
Furthermore, $N=1$, $D=6$ supersymmetry guarantees that there are no higher
($\ge 2$) loop contributions to the gauge field kinetic energy. Positivity
of the kinetic energy for all values of $\phi$ thus implies that $v_\alpha$
and $\tilde v_\alpha$ should both be non-negative, and at least one should
be positive. Otherwise, some interesting new phase transition must occur at
the value of $\phi$ at which the gauge field coupling constant changes sign,
preventing the extrapolation from weak to strong coupling. Since $v_\alpha$
is essentially the Kac-Moody level \cite{Erler,Minasian,evidence} and is
therefore non-negative, the problem devolves upon $\tilde v_\alpha$. 

For $N=2$ heterotic strings obtained by compactification on $T^4$, this is
never a problem because the theory is non-chiral and the Green-Schwarz $d
\tilde H$ vanishes. This can also be seen by noting that this theory is dual
to the Type $IIA$ string compactified on $K3$ \cite{ht} which has no
Chern-Simons terms. For $N=1$ heterotic strings obtained by compactification
on $K3$, however, both $dH$ and $d\tilde H$ are non-vanishing\footnote{This
was the reason for the speculation in \cite{Rahmfeld2} that the dyonic
string might be relevant to heterotic/heterotic duality.} and it is the
rule, rather than the exception, that some of the $\tilde v_{\alpha}$ are
negative! Consider, for example, the SO(32) string with a $k=24$ $SU(2)$
instanton embedded in the $SO(32)$. The resulting $D=6$ gauge group is
$SO(28)\times SU(2)$ and the anomaly eight-form is given by \cite{Erler} 
\be 
I_8 \sim [\tr R^2-\tr F_{SO(28)}{}^2-2\tr F_{SU(2)}{}^2]  
[\tr R^2+2\tr F_{SO(28)}{}^2-44\tr F_{SU(2)}{}^2]\ ,
\la{SO(32)} 
\ee
and the $SO(28)$ coefficient in the second factor enters with the wrong sign
\cite{Minasian,evidence}. A similar problem arises for $E_8 \times E_8$
where one embeds a $k_1$ $SU(2)$ instanton in one $E_8$ and a $k_2=24-k_1$
$SU(2)$ instanton in the other, except in the case of symmetric embedding
where both $k_i=12$. The resulting $D=6$ gauge group is $E_7 \times E_7$ and
for generic embeddings one finds that $v_i=1/6$ but 
\be   
\tilde v_i=\frac{1}{12}(k_i-12)  \ .
\ee
Since we require  $k_1+k_2=24$, one factor will always have the wrong sign
except for the $k=12$ case discussed in \cite{evidence}, for which the
$\tilde v_i$ both vanish.  The anomaly eight-form is thus given by 
\be
I_8\sim [\tr R^2-\frac{1}{6}\tr F_{E_7}^2-\frac{1}{6}\tr 
F_{E_7}^2][\tr R^2] \ .
\la{symmetric}
\ee
Since $\tilde v_\alpha=0$ there is no wrong-sign problem and presumably one
can extrapolate to strong coupling without encountering a phase transition.
Qualitatively similar results hold for any other unbroken subgroup of
$E_8\times E_8$.  Note, however, that since $v_\alpha\not=\tilde v_\alpha$,
there is no manifest self-duality.  In \cite{evidence}, however, the duality
was deduced by looking in two different ways at eleven-dimensional
$M$-theory compactified on $K3 \times S^1/Z_2$. Consequently, one is led to
assume that the duality interchanges perturbative gauge fields
($v_\alpha=0,\tilde v_\alpha<0$), with non-perturbative gauge fields
($\tilde v_\alpha=0,v_\alpha<0$). We shall return to this special case
later, but for the most part the present paper will be concerned with the
generic case where a phase transition seems unavoidable.\footnote{It has
been suggested in \cite{afiq} that a phase transition can also be avoided in
those cases where one can Higgs away the ``wrong-sign'' gauge groups, for
example in the $(k_1,k_2)=(14,10)$ compactification of the $E_8\times E_8$
string.} 

Next, let us recall the {\it dyonic string soliton} \cite{Rahmfeld2} which
carries both ``electric'' charge $Q$ and ``magnetic'' charge $P$. In
canonical metric, it takes the form 
\[
{\phi}={\phi_Q}+{\phi_P}\ ,
\] 
\[
ds^2=e^{({\phi_Q}-{\phi_P})}dx^\mu dx^\nu \eta_{\mu\nu} +
e^{({\phi_P}-{\phi_Q})}dy^mdy^m\ , \]
\[
e^{-2\phi_Q}=e^{-\phi_0}+{Q\over r^2},~~~~~~~e^{2\phi_P}=e^{\phi_0}+{P\over
r^2}\ , \]
\be
H=2P\epsilon_3 + 2Q e^{2\phi}\,{}^*\! \epsilon_3\ , \la{dyon}
\ee
where $x^\mu$ ($\mu=0,1$) are the coordinates of the string world volume,
$y^m$ ($m=1,2,3,4$) are the coordinates of the transverse space,
$r=\sqrt{y^my^m}$ and $\epsilon_3$ is the volume form on $S^3$. Note that we
have taken the metric to be asymptotically Minkowskian, and the asymptotic
value for $\phi$ to be $\phi_0$. The solution describes a single electric
and single magnetic charge at $r=0$ and it interpolates between the purely
electric fundamental string of \cite{Dabholkar} and the purely magnetic
solitonic string of \cite{Lublack,Minasian}. Its tension, or mass $m$ per
unit length, is given by 
\be
2\pi \a'^2\, m=P e^{-\phi_0}+Q e^{\phi_0}\ .
\ee
In the present paper, we are interested in regarding this configuration as a
solution of the chiral $N=1,D=6$ theory describing a graviton multiplet
$(g_{\sst{MN}},\psi_{\sst{M}}, B^+_{\sst{MN}})$ coupled to a single tensor
multiplet $(B^-_{\sst{MN}}, \chi, \phi)$, where the $2$-forms
$B^+_{\sst{MN}}$ and $B^-_{\sst{MN}}$ have $3$-form field strengths that are
self-dual and anti-self dual, respectively. As such, the solution preserves
half of the spacetime supersymmetry for all values of $P$ and
$Q$.\footnote{This is to be contrasted with $N=2$ and $N=4$ theories, where
the purely electric or purely magnetic solutions preserve one half, but the
dyonic solution only one quarter, of the supersymmetry.}  In the self-dual
limit $Pe^{-\phi_0}=Q e^{\phi_0}$, the contributions from the tensor
multiplet become trivial. This self-dual string had already been found in
\cite{Lublack} in the context of self-dual supergravity which describes only
the graviton multiplet and no tensor multiplet. It is also interesting to
consider the limit $Pe^{-\phi_0}=-Qe^{\phi_0}$ where the string becomes
tensionless. We refer to this as the {\it quasi}-anti-self dual limit,
because in this limit the graviton multiplet does not completely decouple,
in that the spacetime is still curved and the self-dual part of the field
strength is still non-vanishing.\footnote{This tensionless string has been
the subject of much discussion recently [15-23] but note that, contrary to
some claims in the literature, it corresponds to the quasi-anti-self-dual
limit of the dyonic string of \cite{Rahmfeld2} and {\it not} the self-dual
string of \cite{Lublack}.  The masslessness of the quasi-anti-self-dual
dyonic string was first observed in \cite{lp}.} 

     We now allow a further coupling to a Yang-Mills multiplet  $(A_{M},
\lambda$) and  turn to the discussion of the {\it gauge dyonic} string
solution of this chiral $N=1$ theory.  The bosonic equations of motion for
the corresponding supergravity are \cite{Sagnotti} 
\bea &&\Box \phi =-\ft1{12} e^{-2\phi} H^2
+\fft{\a'}{16}   \sum_\a (v_\a\, e^{-\phi} - \td v_\a\, e^{\phi}) \tr (F_\a)^2\
,\nonumber\\ &&R_{\sst{MN}} = \del_{\sst M} \phi \del_{\sst N} \phi + \ft14
e^{-2\phi} (H^2_{\sst {MN}} - \ft16 H^2 g_{\sst{MN}}) \nonumber\\
&&\phantom{xxxxxx}-\fft{\a'}{4}  \sum_\a (v_\a\, e^{-\phi} + \td v_\a\,
e^{\phi})\tr (F^2_{\a\sst{MN}} -\ft18 F^2_\a g_{\sst{MN}})\ ,\label{eom1}\\ 
&&D_{\sst{M}}\Big((v_\a\, e^{-\phi} + \td v_\a\,
e^{\phi})F_\a^{\sst{MN}}\Big) -\ft12 v_\a\, e^{-\phi} H^{\sst N}{}_{\sst{PQ}}
F^{\sst{PQ}}_\a - \ft12 \td v_\a\, e^{\phi}\, {}^*\! H^{\sst N}{}_{\sst{PQ}}
F_\a^{\sst{PQ}}=0\ , \nonumber\\
&&d H = \fft{\a'}{4} \sum_\a v_\a\tr\, F_\a\wedge F_\a\ ,\qquad
d \wtd H =  \fft{\a'}{4}\sum_\a \td v_\a\tr\, F_\a\wedge
F_\a\ .\nonumber
\eea
(It is not necessary to include the Lorentz Chern-Simons and Green-Schwarz 
terms in (\ref{Chern}) and (\ref{Green}) since, in common with the gauge 
fivebrane \cite{Strominger}, they will turn out to vanish for the gauge 
dyonic string solution.)

   The ans\"atze for the metric and the field strength $H$ are given by
\bea
ds^2 &=& e^{2A} dx^\mu dx^\nu \eta_{\mu\nu} + e^{-2A} dy^m dy^m\ ,
\nonumber\\
H_{mnp}&=& \epsilon_{mnpq} \del_q e^{C}\ ,\qquad
H_{\mu\nu m}= \epsilon_{\mu\nu} \del_m e^{\wtd C}\ ,\label{ansa}
\eea
The functions $A$, $C$ and $\wtd C$ depend only on $r=\sqrt{y^my^m}$.  Note
that the $\epsilon$ symbols are purely numerical, and the contractions are
performed in the Euclidean metric $\delta_{mn}$.  We shall first consider
the case where the source for the solution is provided by a single $SU(2)$
self-dual Yang-Mills instanton in the 4-dimensional transverse space.  It
can be written as 
\be
F^a = \fft{2\rho^2}{(\rho^2 + r^2)^2} \eta^a_{mn} dy^m\wedge dy^n\ ,
\ee
where $a$ is an adjoint $SU(2)$ index, $\eta^a_{mn}$ are the 't Hooft
symbols, and $\rho$ is the scale size of the instanton.  In order to solve
the equations of motion (\ref{eom1}), we take $C = \phi -2A$ and $\wtd C =
\phi + 2A$. The equations are then all satisfied if $C$ and $\wtd C$ satisfy
\bea
\del_m\del_m e^{C} &=& \fft{\a'}{8}v\, e^{-4A}\, \tr F_{mn}F_{mn}
=-\fft{48\a' v
\rho^4}{(\rho^2 +r^2)^4}\ ,\nonumber\\
\del_m\del_m e^{-\wtd C} &=& \fft{\a'}{8} \td v\, e^{-4A}\, 
\tr F_{mn}F_{mn}=
-\fft{48\a' \td v \rho^4}{(\rho^2 +r^2)^4}\ .\label{eom2} 
\eea
Thus we have
\bea
e^{\phi -2A} &=& e^{C}=e^{\phi_0} + \fft{P(2\rho^2 + r^2)}{(\rho^2 +
r^2)^2} \ ,\nonumber\\
e^{-\phi - 2A} &=& e^{-\wtd C} = e^{-\phi_0} + 
\fft{Q(2\rho^2 + r^2)}{(\rho^2 +r^2)^2} \ ,\label{solution}
\eea
where $Q$ and $P$ are the electric charge and magnetic charge, given by
\be
Q\equiv \ft1{2\omega_3} \int_{S^3} \wtd H =2\a' \td v\ ,\qquad P
\equiv \ft1{2\omega_3} \int_{S^3} H=2\a' v\ .
\ee
Here $\omega_3$ denotes the volume of the unit three sphere $S^3$.  The mass
per unit length of this dyonic string is given by 
\be
2\pi \a'^2\, m= Pe^{-\phi_0} + Qe^{\phi_0}\ .\label{mass}
\ee
Note that since $\del_m\del_m r^{-2}=0$, we could in principle add $r^{-2}$
terms with arbitrary coefficients to the solutions for $e^{C}$ and $e^{-\wtd
C}$. However, such terms would describe contributions to the string mass and
charges coming from singular sources rather than from the instanton source
that we are considering here.  As one might expect, we recover the neutral
dyonic solution (\ref{dyon}) as we shrink the size of the instanton to zero.

    We may count the bosonic and fermionic zero modes by following the same
procedure used for the gauge fivebrane \cite{Strominger}.  For concreteness
we consider the case where the $SU(2)$ is embedded minimally in an $E_7$
gauge group, {\it i.e.}\ $E_7 \longrightarrow SO(12) \times SU(2)$.  First
we count the bosonic modes: There will be 4 translations, 1 dilatation, 3
$SU(2)$ gauge rotations and 64 modes coming from the embedding, given by the
dimension of the coset $E_7/(SO(12)\times SU(2))$.  The counting of the
fermionic zero modes is provided by the index theorem.  Bearing in mind that
the trace of $F\wedge F$ in the adjoint of $SU(2)$ is 4 times the trace in
the doublet, and noting that the dual Coxeter number of $E_7$ is 18, we find
$4 \times 18=72$ fermionic zero modes.  Thus we have a total of $72+72$
bosonic and fermionic zero modes, which presumably corresponds to a
non-critical string.   Qualitatively similar results apply for other choices
of gauge group. 

    This construction can be easily extended to the case where the source is
provided by an $SU(2)$ multi-instanton configuration.  This is particularly
simple for the 't Hooft \cite{th} and the Jackiw, Nohl and Rebbi \cite{jnr}
classes of multi-instanton solutions.  In these solutions, $\tr F_{mn}F_{mn}
= 4\del^2\, \del^2 \log f$, where $f=\epsilon + \sum_i \mu_i|\vec y - \vec
y_i|^{-2}$, with $\epsilon=1$ or 0 respectively.  Thus the solutions for $C$
and $\wtd C$ are given by 
\be
e^{C} =\ft12 \a' v (\del_m\del_m f + h)\ ,\qquad
e^{-\wtd C} = \ft12 \a' \td v (\del_m\del_m f + h)\ ,
\ee
where $h$ is a solution of the homogeneous equation $\del_m\del_m h = 0$,
chosen so that the dyonic string solution has no other sources than those
provided by the multi-instanton configuration.  Note that these
configurations break the isotropicity of the transverse space, and thus the
functions $C$, $\wtd C$ and hence $A$ and $\phi$ depend non-isotropically on
the coordinates $y^m$.  The string solution has an electric charge and a
magnetic charge, which are given by $Q=2\a' n \td v$ and $P= 2\a' n v$
respectively, where $n$ is the instanton number.  The mass $m$ per unit length
is still given in terms of these charges by (\ref{mass}). If $SU(2)$
instantons in more than one of the components in the original Yang-Mills group
are considered, the charges will be given by 
\be
Q = 2\a'\sum_\a n_\a \td v_\a\ ,\qquad P = 2\a' \sum_\a 
n_\a v_\a\ ,\label{charge}
\ee 
where $n_\a$ is the instanton number for the $\a$'th component of the
Yang-Mills group. 

     Let us now consider the supersymmetry of the gauge dyonic string
solutions.  The transformation rules for the fermionic fields are 
\bea
\delta \psi_{\sst M} &=& D_{\sst M}\epsilon + \ft1{48} e^{-\phi} H^{\sst
{NPQ}} \Gamma_{\sst{NPQ}} \Gamma_{\sst M} \epsilon\ ,\nonumber\\ 
\delta \chi &=& \ft{\rm i}2 \del_{\sst M} \phi \Gamma^{\sst M} \epsilon
+\ft{\rm i}{12} e^{-\phi} H_{\sst{MNP}} \Gamma^{\sst{MNP}} \epsilon\ ,
\label{susy}\\
\delta \lambda &=& -\ft{1}{2\sqrt2} F_{\sst{MN}} 
\Gamma^{\sst{MN}}\epsilon\ .\nonumber
\eea
It is straightforward to substitute the gauge dyonic string solutions into
these transformation rules, and we find that the variations of the fermion
fields vanish if 
\be
\epsilon = e^{\ft12 A} \epsilon_0\ ,\qquad
\Gamma_{01} \epsilon_0= \epsilon_0\ ,
\ee
where $\epsilon_0$ is a constant spinor.  Thus the gauge dyonic soliton also
preserves half of the $N=1$, $D=6$ supersymmetry for all non-vanishing
values of the electric charge $Q$ or the magnetic charge $P$.  Note that 
supersymmetry requires that the Yang-Mills configuration in the transverse 
space be self-dual, and therefore the instanton numbers must all be 
non-negative.  If $Pe^{-\phi_0}=Q e^{\phi_0}$, the solution reduces to the
gauge self-dual string, where the anti-self-dual component of $H$ vanishes
and the dilaton becomes constant, given by $\phi=\phi_0$, and hence the
complete tensor multiplet decouples from the theory.  On the other hand, if
$Pe^{-\phi_0}=-Qe^{\phi_0}$ the string soliton becomes massless and
quasi-anti-self-dual. 

      Having obtained the generic gauge dyonic string solutions, their
connection with the phase transition now becomes apparent.  As we stated
earlier, it follows from (\ref{happiness}) that positivity of the kinetic
energy for all values of $\phi_0$ requires that $v_\a$ and $\td v_\a$ should
be non-negative, and at least one should be positive.  In these vacua, the
gauge dyonic string soliton has non-vanishing positive mass.  It is in those
string vacua where one or more of the $\td v_\a$ is negative that the
possibility of the tensionless string arises, as can be seen from
(\ref{mass}) and (\ref{charge}).  As a concrete example, let us consider the 
case where just one of the $\td v_\a$ coefficients is negative, say $\td
v_1$. Thus the phase transition takes place at the point $\phi_0= \phi^{\rm
cr}_0$, where 
\be
v_1 e^{-\phi_0^{\rm cr}} + \td v_1 e^{\phi_0^{\rm cr}}=0\ .\label{critphi}
\ee
The tension, on the other hand, is given by
\be
\pi \a'\, m = n_1 (v_1 e^{-\phi_0} + \td v_1 e^{\phi_0}) +
\sum_{\a\ge 2} n_\a (v_\a e^{-\phi_0} + \td v_\a e^{\phi_0} )\ .
\ee
Since the quantities in the summation over $\a \ge 2$ are all non-negative,
the tension will be positive for $\phi_0 < \phi_0^{\rm cr}$.  At the phase 
transition, the string tension remains positive for generic values of the
instanton numbers, but becomes zero if $n_\a =0$ for $\a\ge 2$. (There are 
further tensionless strings for other choices of instanton numbers if the
resulting electric charge $Q=2\a'\sum_{\a\ge 1} \td v_\a$ is negative.
These all occur at larger values of the coupling, {\it i.e.}\ $\phi_0>
\phi^{\rm cr}_0$.  With any further increase of the coupling, the tension of
a tensionless string becomes negative.   It should be emphasised that
neither of these phenomena can occur in the $\phi_0 < \phi^{\rm cr}_0$
regime where the formalism can be trusted.) 

      The above discussion can be easily generalised to the cases where more 
than one $\td v_\a$ is negative.   The phase transition occurs at the 
smallest value of the string coupling at which the coefficient of any of 
the Yang-Mills kinetic terms becomes zero. This value of the string coupling
$e^{\phi_0}$ at which the phase transition takes place is always accompanied
by a massless string.  This phenomenon is reminiscent of the conifold
transitions in Calabi-Yau compactification of the type II string induced by
massless black holes \cite{str3}.  However, in our case, for the larger
values of the coupling when the kinetic energy of the gauge field
$F_{\a\sst{MN}}$ becomes negative, the mass of the dyonic string also
becomes negative.  As in the case where only one of the $\td v_\a$
coefficients is negative, here also there exists an infinite subset of the
other choices of instanton number for which the value of $e^{\phi_0}$ where
the string becomes massless does not correspond to the phase transition; in
fact these tensionless dyonic strings all occur at larger values of the
coupling $e^{\phi_0}$.  Thus all the string solutions have positive definite
tension in the weak coupling regime before the onset of the phase
transition, whilst all the tensionless solutions occur at or beyond the
phase transition, and all the negative-tension string solutions occur beyond
the phase transition. 
     
      Now we turn to the issue of naked singularities of the string
solution.  We can study this by looking at the scalar curvature.  For the
6-dimensional metric $ds^2 = e^{2A} dx^\mu dx^\nu \eta_{\mu\nu} + e^{2B}
dy^mdy^m$, it is given by $R = -e^{-2B}( e^{A-B}\, \del_m\del_m e^{B-A} +
5e^{-A-B}\,\del_m\del_m e^{A+B})$.  It is easy to verify, for the canonical
metric $g_{\sst{MN}}$ (\ref{ansa}) and for the metrics $e^{\pm \phi}
g_{\sst{MN}}$ of the string and the dual string, that the scalar curvature
diverges if either $e^C$ or $e^{-\wtd C}$, given by (\ref{solution}),
vanishes.  If this occurs for a positive value of $r^2$, the metric will 
have a naked singularity.  We see from (\ref{solution}) and (\ref{charge})
that in order to avoid a naked singularity for all possible instanton
numbers $n_\a$ and all possible values of the instanton sizes $\rho_\a$, the
coefficients $v_\a$ and $\td v_\a$ should both be non-negative. This is
precisely the requirement that the theory not undergo a phase transition! We
conclude, in particular, that naked singularities never arise in those vacua
admitting a heterotic/heterotic self-duality. 
         
     As the present paper was nearing completion, we became aware of the
paper by Seiberg and Witten \cite{sw}, which also relates phase transitions
to tensionless gauge dyonic strings.

%%%\bibliographystyle{preprint} %%%\bibliography{duality}
%%%%%%%%%%%%% include bibtex generated bibliography %%%%%%%%%%%%%%%

%%%%%%%%%%%%%%%%%% end bibliography %%%%%%%%%%%%%%%%%%%%%%%%%%%%%%%

\end{document}